%% file: main.tex
\documentclass[10pt,letterpaper,compsoc,conference]{iiswc24}

\usepackage{cite}
\usepackage{amsmath,amssymb,amsfonts}
\usepackage{algorithmic}
\usepackage{graphicx}
\usepackage[dvipsnames]{xcolor}
\usepackage[final]{microtype}
\usepackage[italic]{mathastext}
\usepackage{libertine}
\usepackage[T1]{fontenc}
\usepackage{textcomp}
\usepackage[varqu,varl]{zi4}
\usepackage[all]{nowidow}
\usepackage[auth-lg,affil-it]{authblk}
\usepackage[keeplastbox]{flushend}
\usepackage{fancyhdr}
\usepackage{enumitem}

\usepackage{tabularx}
\usepackage[most]{tcolorbox}
\usepackage{multicol}
\usepackage{multirow}
\usepackage[dvipsnames]{xcolor}

\tcbset{
    frame code={}
    center title,
    left=0pt,
    right=0pt,
    top=0pt,
    bottom=0pt,
    colback=gray!26,
    colframe=white,
    width=\linewidth,
    enlarge left by=-6mm,
    enlarge right by=-3mm,
    boxsep=5pt,
    arc=0pt,outer arc=0pt,
    nobeforeafter,
    before={\vspace{0.05cm}}
}

\begin{document}


\title{Evaluating the Effectiveness of Microarchitectural Hardware Fault Detection for Application-Specific Requirements} 


\renewcommand\Authsep{\qquad}
\renewcommand\Authand{\qquad}
\renewcommand\Authands{\qquad}



\author[1]{Konstantinos-Nikolaos Papadopoulos}
\author[2]{Christina Giannoula}
\author[1]{Nikolaos-Charalampos Papadopoulos}
\author[1]{\\Nektarios Koziris}
\author[3]{José M.G. Merayo}
\author[1]{Dionisios N. Pnevmatikatos}
\affil[1]{National Technical University of Athens}
\affil[2]{University of Toronto}
\affil[3]{Technical University of Denmark}


\maketitle
\pagestyle{plain}


\input{Abstract}
\input{Introduction}
\input{Related}
\input{Motivation}
\input{Background}

\input{Methodology}
\input{Evaluation}
\input{OtherRelated}

\input{Conclusion}

\bibliographystyle{IEEEtranS}
\bibliography{References}

\end{document}

%% file: Abstract.tex
\begin{abstract}
Reliability is necessary in safety-critical applications spanning numerous domains. Conventional hardware-based fault tolerance techniques, such as component redundancy, ensure reliability, typically at the expense of significantly increased power consumption, and almost double (or more) hardware area. To mitigate these costs, microarchitectural fault tolerance methods try to lower overheads by leveraging microarchitectural insights, but prior evaluations focus primarily on only application performance. As different safety-critical applications prioritize different requirements beyond reliability, evaluating only limited metrics cannot guarantee that microarchitectural methods are practical and usable for all different application scenarios. 
To this end, in this work, we extensively characterize and compare three fault detection methods, each representing a different major fault detection category, considering real requirements from diverse application settings and employing various important metrics such as design area, power, performance overheads and latency in detection. Through this analysis, we provide important insights which may guide designers in applying the most effective fault tolerance method tailored to specific needs, advancing the overall understanding and development of robust computing systems. 
For this, we study three methods for hardware error detection within a processor, i.e., (i) Dual Modular Redundancy (DMR) as a conventional method, and (ii) Redundant Multithreading (R-SMT) and (iii) Parallel Error Detection (ParDet) as microarchitecture-level methods. We demonstrate that microarchitectural fault tolerance, i.e., R-SMT and ParDet, is comparably robust compared to conventional approaches (DMR), however, still exhibits unappealing trade-offs for specific real-world use cases, thus precluding their usage in certain application scenarios.
\end{abstract}


%% file: Introduction.tex
\section{Introduction}

Modern computing systems demand strong robustness, safety, and/or security properties, thus requiring high tolerance against transient and permanent hardware faults, such that to prevent failures and guarantee uninterrupted and reliable operation \cite{cappello2009toward, harddata, reliable_nanoera, SEU_ground, alibaba_sdc}. Hardware failures can jeopardize human lives, or disrupt the successful completion of critical and costly tasks \cite{Australian_report, NASA_Toyota_report, ISO26262, do254}. For microelectronic circuits operating in high radiation environments, increased reliability is paramount due to radiation interference, which causes highly increased error rates \cite{Single_event_upset_space}. However, errors are present even in lower radiation environments. This is because technology scaling in modern Integrated Circuits (ICs), aimed at enhancing performance and energy efficiency through the miniaturization of transistor sizes, also increases susceptibility to radiation-particle induced faults \cite{technology_scaling_reliability, CMOS_Miniaturization_Cosmic-Ray}. Moreover, this susceptibility is expected to increase further with continued scaling. Likewise, chip manufacturers increase reliability by introducing conservative margins in operating voltage and frequency, and future microprocessors will need robust tolerance to transient errors in order to exploit lower margins and harness back power and performance gains \cite{micro_voltage_scaling_papadim, adaptive_guardband}.   

All such systems require fault tolerance mechanisms, like fault detection, which however, in order to increase reliability come with the cost of either negatively impacting performance, or increasing power consumption, or IC area, depending on the method design and trade-offs. As a result, different methods exhibit different overheads in a number of metrics and represent distinct points in the fault detection method design space. The selection of the most suitable method, depends on the constraints of the design and the application domain, since apart from reliability, critical systems from various application domains have different requirements depending on their particular characteristics. For example, High Performance Computing (HPC) applications, which have growing needs for reliability due to scaling and tight margins, but they also primarily need high performance to solve complex problems and process massive amounts of data \cite{dongarra2011international, hpc-survey}. Instead, deep space missions prioritize the need for low area and power costs due to energy constraints in spacecraft probes with limited battery capacities \cite{psyche, parker}. Therefore, in both cases hardware fault tolerance mechanisms have to be deployed, which however need to meet different constraints: the first mechanism must exhibit low performance overheads, while the second one low area and power overheads. In like manner, other application scenarios have other distinct requirements for performance, area or power and demand fault tolerance mechanisms which comply with their unique constraints.
As a result, engineers need to have an accurate estimation on the strengths and weaknesses of each method, covering various metrics apart from performance, when choosing one for a particular application. 

\begin{figure}
\centerline{\includegraphics[width=\linewidth]{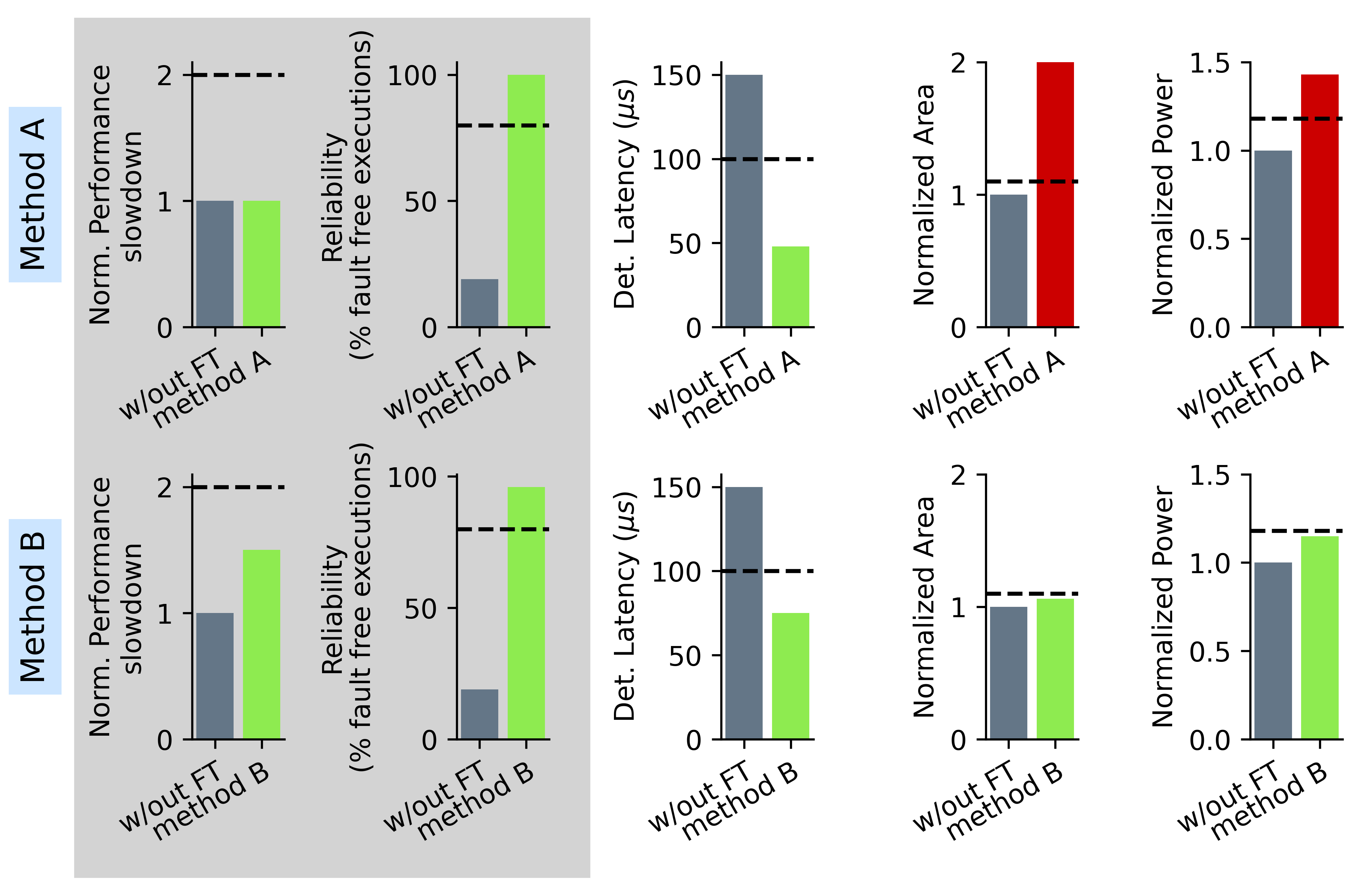}}
\caption{
Different fault detection mechanisms can exhibit similar performance and reliability but differ significantly in other unevaluated metrics (detection latency and area). Thus, evaluations must consider all relevant metrics. Here, Method A (DMR) and Method B (R-SMT) are assessed against design constraints from a nano-satellite application. Green bars meet the constraints, while red bars violate them.}
\label{fig:motivation-sch}
\end{figure}


Prior works propose effective error detection methods which leverage microarchitectural insights \cite{diva1, diva2, restore, ainsworth, reliabikity-runahead, arsmt, fault-det-via-smt, parashar_slick:_2006, gomaa_opportunistic_2005, DTE, mukherjee_detailed_2002}, aiming to achieve high performance and low area footprint of the secured systems, and reduce by-design some of the overheads associated with fault tolerance. However, such works mainly focus on the performance aspect of their methods, and do not comprehensively evaluate overheads in other metrics, such as area or power (i.e. \cite{diva1, gomaa_opportunistic_2005,DTE,mukherjee_detailed_2002,reliabikity-runahead,parashar_slick:_2006,arsmt,sundaramoorthy_slipstream_2000,restore}), which are of crucial importance for some application domains.
In Fig. \ref{fig:motivation-sch}, we demonstrate that by assessing only performance and reliability, evaluations of microarchitectural fault detection methods are not complete nor adequate: as we find after reviewing a wide range of safety-critical applications (Section \ref{sec:key_reqs}), many such systems demand low overheads in other metrics, like area, power or detection latency, rather than performance.
Thus, prior works mainly cover a subset of the use case scenarios that need high performance guarantees (e.g., HPC applications) along with reliability, and overlook other use cases, such as space missions or automotive, that have different priorities. 

Fig. \ref{fig:motivation-sch} evaluates various important metrics and shows the overheads of two different fault detection methods (Method A and Method B) compared to without any fault detection ("w/out FT", in grey bars). Fig. \ref{fig:motivation-sch} also shows the constraints of a nano-satellite system \cite{tadanki2019closing, clark2011power} that requires fault detection with the dashed horizontal lines. When choosing between the two candidate fault detection mechanisms (Method A and B), if engineers only consider existing evaluations focused on performance and reliability (grey background plots), both methods appear to meet these application constraints, and Method A might be selected. However, when additional metrics critical for orbital systems, such as detection latency, IC area and power, are assessed, it becomes evident that Method A is unsuitable due to its high area and power overheads, which cause the total area cost of the system to exceed the area and power constraints (illustrated with a red bar). Instead, Method B meets these constraints of the application, thereby being the most suitable method to be selected. 
This example underscores the need for a comprehensive evaluation of fault detection methods across a variety of important metrics tailored to assess the requirements of target applications.

To this end, the \emph{goal} of our work is to extensively characterize a diverse set of different fault detection methods in various key requirements. To our knowledge, there is no prior work to extensively compare and examine microarchitectural fault detection methods across a wide range of use case application scenarios that have different key requirements. Therefore, we conduct the following: 

Firstly, we review various application domains requiring robustness properties -both safety-critical and emerging ones (like HPC)- and identify that they have distinct yet important requirements beyond reliability, which however, are overlooked in most evaluations of prior hardware fault tolerance mechanisms. Based on these requirements, we group application domains on real-world use-cases to three types: (a) performance-critical, as the applications which need high performance guarantees, (b) area/power-critical, as the applications that need low area and power overheads, and (c) latency-critical, as the applications which need low error detection latency. 

Secondly, based on these use-case categories, we propose a better evaluation methodology that more appropriately compares fault detection methods and their trade-offs. We provide the key metrics on which hardware error detection needs to be evaluated in order to understand in which real-world applications is best fit and in which ones might result in important application degradation. These metrics consist of (i) detection efficiency -namely the effectiveness in error detection- (ii) detection latency -the elapsed time between fault manifestation and error detection- (iii) performance, (iv) design area and (v) power. 

Thirdly, we analyse prior works on hardware error detection and select 3 diverse redundancy approaches across both microarchitectural and conventional methods: (i) hardware redundancy, (ii) multithreading within microprocessors, and (iii) heterogeneous error detection mechanisms. This selection captures different fault detection capabilities, providing valuable insights into their effectiveness and efficiency in maintaining system reliability. We further select one method from each of these three major fault detection method classes: We choose Dual Modular Redundancy, a conventional method of hardware redundancy, Redundant Multithreading via Simultaneous Multithreading (R-SMT), a microarchitectural method which utilizes multithreading within microprocessors and is representative of the design space encompassing all prior redundant multithreading techniques, and the most recent state-of-the-art heterogeneous fault detection method, Parallel Error Detection with Heterogeneous Cores (ParDet). DMR, is a method of hardware redundancy that 
duplicates whole hardware components, providing robustness at the expense of significantly increased power consumption and design area. R-SMT \cite{arsmt, fault-det-via-smt, parashar_slick:_2006, gomaa_opportunistic_2005, DTE, mukherjee_detailed_2002} duplicates the execution in redundant hardware threads. ParDet \cite{ainsworth} uses multiple parallel heterogeneous cores to re-execute parts of the program independently. Please refer to Section \ref{sec:prior_work} for a comprehensive description of all prior fault detection methods.

Fourth, we implement and evaluate all methods using the gem5 simulator \cite{gem5}, in same core configuration for fair comparison. We implement statistical fault injection to induce transient and permanent faults to all methods, covering the whole range of appeared faults in the reviewed application domains. 
This approach allows to assess reliability in the microarchitectural level, offering both accurate and early in the design stage reliability results. 
We  evaluate all representative methods using our proposed evaluation methodology (of 5 metrics) and a variety of workloads. 

We provide new insights and important key observations that can be leveraged to find which method is the most appropriate for a particular application-specific computing system. By comparing DMR, R-SMT, and ParDet, researchers can gain a holistic view of the trade-offs involved in different microarchitectural redundancy strategies, their impact on system robustness, and their suitability for various applications. This comparative analysis helps in identifying the most effective fault tolerance method tailored to specific needs and advances the overall understanding and development of robust computing systems. 

Our most significant key insights are:
\begin{itemize}
    \item Microarchitectural methods i.e., R-SMT and ParDet, can have comparably high capability on identifying errors compared to conventional methods, i.e., DMR.
    \item Low detection latency is crucial in identifying permanent errors to guarantee high detection efficiency.
    \item R-SMT is suitable for area-critical applications.
    \item ParDet is suitable for performance-critical applications thanks to providing best performance across all evaluated workloads.
\end{itemize}

Overall, we make the following key contributions: 
\begin{itemize}
    \item We find that fault detection methods have different trade-offs in reliability, latency, performance, area and power, and that different applications have different priorities (apart from the reliability which is always the first) among these requirements. We group applications in three categories based on their secondary priority and this categorization can assist both when designing new detection methods to navigate the requirements of the prospective target applications and in selecting suitable existing methods for a specific domain.
    
    \item We propose a more appropriate evaluation methodology for fault detection methods, which consists of five key metrics. We believe that our evaluation methodology will be adopted by our research community and industry, when proposing new fault detection methods, in order for new proposed mechanisms to be proven practical and usable in the targeted application types.
    
    \item We implement and comprehensively evaluate the most representative hardware fault detection methods from a wide range of fault detection strategies and across all crucial metrics, using the methodology we establish. The new insights we provide will be useful to industry engineers for selecting the optimal methods in different application domains when designing application-specific computing systems and enhancing them with the corresponding fault detection methods.
    
\end{itemize}

%% file: Related.tex
\section{Hardware Error Detection Methods} \label{sec:prior_work}

We choose to evaluate one method from each different category of hardware fault detection, namely Hardware Redundancy, Redundant Multithreading and Heterogeneous Systems, to provide insights for a variety of different fault detection strategies.

\subsection{Hardware Redundancy} 
Spacial hardware redundancy, where the whole core or individual components are duplicated and the outputs of both copies compared for fault detection, has been used in many commercial systems like the IBM G5 \cite{ibms490}, Tandem (now HP) NonStop \cite{tale2systems, nonstop}, and others \cite{teramac}, most recently in ARM Cortex-R CPUs \cite{cortexR, cortexRassesment}. Component redundancy, is the primary method of choice in high-risk space missions, due to the high reliability it provides \cite{space_mission_analysis, silva2020cevero}. 
Additionally, multi-core architectures like Chip Multi-Processors (CMPs) have also been proposed as a platform for DMR \cite{resilientCMP, CMP_dmr_tmr, stagenet, mukherjee_detailed_2002}. In this study we evaluate a DMR scheme with one redundant core, since it provides increased robustness compared to individual component redundancy and to also provide a baseline of conventional fault detection.

\subsection{Redundant Multithreading}
Redundant Multithreading employs different hardware threads to run the same program, and compares their results for fault detection. Rotenberg \cite{arsmt} exploits two different instruction streams (A-stream and R-stream) which redundantly run on the same processor using Simultaneous Multithreading for detecting faults. Mukherjee and Reinhardt\cite{fault-det-via-smt} expand this, evaluating performance improvements like branch and operand prediction by the leading thread and propose the idea of sphere-of-replication. 

Other approaches \cite{parashar_slick:_2006, gomaa_opportunistic_2005, sundaramoorthy_slipstream_2000} aim at reducing the execution overhead of the redundant thread, introducing partial redundant multithreading: SlicK \cite{parashar_slick:_2006} re-executes only store instructions along with their backward slices, eliminating completely slices which are predicted to be fault-free from the redundant thread. Opportunistic Transient Fault Detection \cite{gomaa_opportunistic_2005} executes a redundant thread during low ILP phases and cache misses and for high ILP phases exploits instruction reuse to reduce the re-execution of the redundant thread. Slipstream processors \cite{sundaramoorthy_slipstream_2000} implement a pair of streams, assisting each other in both performance and fault tolerance, with the A-stream running ahead of R-stream and eliminating unnecessary instructions from the R-stream. All these works target to improve performance over the initial design \cite{arsmt, fault-det-via-smt}, while maintaining reliability, degrading however other metrics, which as we identify in Section \ref{sec:key_reqs} are of equal or even greater importance, depending on the application setting. 

Additionally, prior works only evaluate their microarchitectural methods in terms of detection efficiency and performance, thus not covering all key metrics needed to comprehensively assess the methods in real-world application scenarios. In this study, we aim to compare methods between the different categories to provide general insights for a variety of different hardware strategies, therefore as a representative method from the Redundant Multithreading category, we evaluate the original design. Other variations (i.e. \cite{parashar_slick:_2006, gomaa_opportunistic_2005, sundaramoorthy_slipstream_2000} are variations of either full or partial Redundant Multithreading and therefore represent close points in the design space. 

However, several prior works \cite{parashar_slick:_2006,gomaa_opportunistic_2005, DTE,mukherjee_detailed_2002} propose performance optimizations upon the original R-SMT scheme. In our work, we choose not to include additional performance optimizations in the R-SMT scheme. Such optimizations like perfect branch prediction and operands prediction \cite{arsmt} provide slight performance improvements over R-SMT (without affecting reliability), but negatively affect other metrics such as area overheads. These inherit trade-offs of the mechanism, do not drastically alter its position in the design space; all R-SMT variations represent close points, improving some metrics but deteriorating others. Since we want to provide insights for a variety of different redundancy strategies, we evaluate the lower-bound for R-SMT scheme, without any performance optimizations.  

\subsection{Heterogeneous Systems} 
Austin et al. \cite{diva1, diva2} combines the main core with a single checker core to re-execute instructions upon commit enabling both error detection and correction. However, this mechanism leaves the pipeline frontend unprotected, thus incurring significantly lower detection coverage. Necromancer \cite{necromancer} mitigates permanent manufacturing faults in CMPs. In this scheme, special, lighter, cores are added in the CMP and defective cores continue executing, providing performance enhancing hints to the lighter ones, which substitute the defective ones. Lastly, Parallel Error Detection with Heterogeneous Cores (ParDet) \cite{ainsworth} is the state-of-the-art method in Heterogeneous Fault Detection methods, which utilizes small cores for error detection and exhibits low overheads in performance, area and power. In our analysis, we evaluate ParDet, as the latest and mostly performant efficient heterogeneous fault detection method.

%% file: Motivation.tex
\section{Key Requirements of Safety-Critical Applications} \label{sec:key_reqs}

Tolerating as many hardware errors as possible to provide continuous safe and reliable operations, is a prerequisite in safety-critical systems including space \cite{space_mission_analysis}, automotive \cite{ft-automotive}, healthcare \cite{Pacemaker_reliability}, and nuclear safety systems \cite{osti_1238384}. However, in such systems, robustness is not \textit{the sole} requirement. In addition to robustness -which is always the first priority for all critical systems- different application scenarios might need to prioritize different requirements as a second priority. We analyze several critical application scenarios, and find that they can be classified in three different categories, depending on their second-level priority requirement:

\noindent \textbf{(1) Performance-Critical:} Safety-critical applications that need to provide high-performance capabilities along reliability can be grouped as \textit{performance-critical}. For instance, HPC workloads target high performance efficiency, while they need to tolerate increased error rates due to transistor scaling. Certain automotive use cases also demand high performance \cite{road_automotive_performance}, fueled by the heavy computational requirements of autonomous driving. Similarly, emerging applications in spaceborne systems need to deliver even more high performance \cite{sat_visions_challenges} while retaining high fault tolerance, driven by the integration of artificial intelligence \cite{ai_sat_opportunities, Space-air-ground-survey} and edge computing \cite{Orbital_Edge_Computing, space_Microdatacenters}, which require more compute capabilities.

\noindent \textbf{(2) Area/Power-Critical:} In many important applications, the first priority -additionally to fault tolerance- is the minimization of silicon area, and hence such applications can be considered as \textit{area/power-critical}. These include energy-constrained systems, such as on-board computers in deep-space missions \cite{Le1999MiniaturizationOS, ultra-thin-space, sparrow-date}: larger IC area corresponds to increased energy consumption and increased probability of radiation particle collisions, which thus significantly increase the probability of errors. 

\noindent \textbf{(3) Latency-Critical:} Latency in error detection, i.e., the time period from the time that the error arises to the time that the error is detected by the method, is another key requirement that needs to be optimized \cite{Safety_critical_systems} in many different application domains (space \cite{butler_primer_2008}, automotive \cite{automotive_sae_latency}, nuclear \cite{ft_nuclear}), forming the class of \textit{latency-critical} applications. Increased latency not only prolongs the response time of error-hardened systems, but also increases the performance overheads in error corrections (late detection of an error requires replaying more instructions to restore a safe state) and compromises reliability (accumulated faults that have not been yet detected significantly increase the failure risk, which is particularly important in systems with limited error tolerance).

Overall, different critical application scenarios have different second-level needs, i.e., they need to prioritize either performance, area, and latency, second to reliability. Fault detection mechanisms impose different performance and area overheads, and also have different degrees of detection efficiency and latency, thus influencing all the above requirements. Prior works \cite{diva1, gomaa_opportunistic_2005,DTE,mukherjee_detailed_2002,reliabikity-runahead,parashar_slick:_2006,arsmt,sundaramoorthy_slipstream_2000,restore} mainly focus their evaluations of existing fault detection methods only on performance metrics, thus targeting only a subset of the applications (e.g., performance-critical such as the HPC focused only). To this end, we conduct a comprehensive evaluation study of the a diverse set of microarchitecture-level fault detection methods to present, for the first time, meaningful insights and trade-offs that cover all the aforementioned application requirements. To facilitate such a complete comparative evaluation, we consider the following metrics:
\begin{enumerate}[noitemsep,topsep=0pt,nolistsep,leftmargin=12pt,nosep]
    \item \noindent\textbf{detection efficiency:} the effectiveness in accurately detecting the errors occurred.
    \item \noindent\textbf{detection latency:} the time difference between error manifestation and its detection. 
    \item \noindent\textbf{performance overhead:} the system performance degradation imposed by the error detection method.
    \item \noindent\textbf{area overhead:} the additional surface area required for the error detection resources.
    \item \noindent\textbf{power overhead:} the additional power consumption due to the implementation and function of the error detection mechanism.
\end{enumerate}


%% file: Background.tex
\section{Description of Evaluated Methods}


In this section, we present the main functionality and characteristics of each method that is considered in this paper. All three methods enhance the microprocessor’s reliability in the presence of transient and permanent faults.

\subsection{Spatial Dual Modular Redundancy (DMR)}
DMR is a spatial redundancy technique, in which each hardware component is replicated and computation is repeated in both component copies of the system, to provide reliable operation through redundancy. We evaluate a DMR scheme that consists of two single core processors which share the same memory subsystem. Both cores are executing the same instructions concurrently. As shown in Fig. \ref{fig:methods-sch}a, each instruction of the main core (green segments) is executed concurrently with the corresponding same instruction of the redundant core (yellow segments). In this scheme, both transient and permanent errors are detected by comparing the instruction results from the two processors. 

\subsection{Redundant Simultaneous Multithreading (R-SMT)}
\begin{figure}
\centerline{\includegraphics[width=\linewidth]{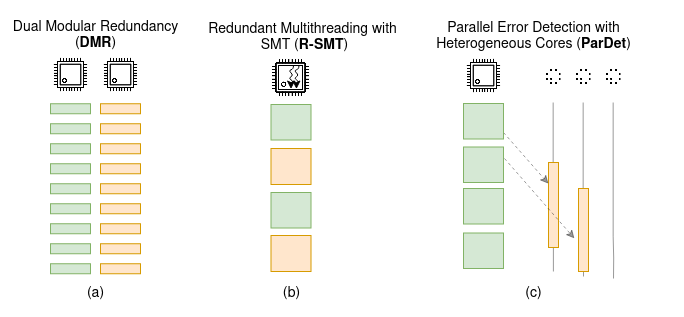}}
\caption{High-level overview of state-of-the-art hardware detection methods. Green segments represent the main execution, while yellow ones the redundant execution for error detection.}
\label{fig:methods-sch}
\end{figure}

R-SMT \cite{arsmt, fault-det-via-smt, parashar_slick:_2006, gomaa_opportunistic_2005, DTE, mukherjee_detailed_2002}, is a class of time redundancy fault-tolerant technique, in which redundant execution is taking place into two different SMT threads running on the same processor core, and redundantly executing the same instructions. Fig. \ref{fig:methods-sch}b shows the main and redundant instructions from the two threads interleaving execution within the core. Then, the two SMT threads validate the instruction results for error detection. 
To do so, instruction results of the primary thread are stored to a hardware buffer (named \textit{comparison buffer}) and consumed by the redundant thread which subsequently utilizes them to perform result comparisons. When the comparison buffer is full, the primary thread stalls (\textit{full comparison buffer stalls}), until redundant thread consumes some entries. Similarly, when the buffer is empty, the redundant thread stalls, until it gets filled again.

\subsection{Parallel Error Detection with Heterogenous Cores (ParDet)} 
ParDet \cite{ainsworth} is a microarchitectural method to retain low area, power, and performance overheads. ParDet parallelizes fault detection by combining the main processor with auxiliary low power processor cores, that redundantly repeat the same instructions with the main core. 

Specifically, the execution on the main core is segmented into parts, each consisting of consecutive instructions. Each segmented part after being executed on the main core is offloaded to one auxiliary (checker) core for re-execution (as shown in Fig. \ref{fig:methods-sch}c), along with the architectural state before (starting state) and after (ending state) that part's execution. Each checker core is initialized with the starting architectural state and executes the current segment. If no errors have occurred after the execution of the segment’s instructions, its architectural state is expected match the provided ending state, so errors are detected by comparing these two architectural states. This way, program segments can be verified independently and in parallel across multiple low-power auxiliary cores (as illustrated in Fig. \ref{fig:methods-sch}c with the overlap of yellow segments in the y-axis, which represents time). The process of copying the starting and ending architectural states is called \textit{checkpointing}. Finally, all values that are read and written by the main core are duplicated in a hardware buffer (\textit{load-store log}), which checker cores can access. The load-store log is split in segments, the number of which is equal to the number of checker cores. Offloading to a checker core is performed, when the segment of the load-store log that corresponds to that checker core is completely filled, or after a certain number of instructions. 


%% file: Methodology.tex
\section{Methodology}
We evaluate the three methods in the gem5 simulator \cite{gem5}, implementing DMR and R-SMT and using the original artifact \cite{ainsworth_pardet_artifact} for ParDet. We choose a representative set of 9 benchmarks from MiBench suite \cite{mibench}. MiBench has been widely used by prior fault tolerance and reliability assessment studies \cite{fiasco, papadim_TC23, papadim_hpca23, papadim_isca21, Demystifying}, as it consists of realistic workloads with reasonable execution time, therefore enabling the thousands of executions required for such a comprehensive fault injection experiment, to complete within reasonable simulation time. Additionally, the workloads we choose span a wide variety of different application domains, as presented in Table \ref{tab:bench}, suitable for our application-specific analysis.

\input{tables/benchmarks} %

Table \ref{tab:config} summarizes the system configuration we use in our evaluation experiments. We model a medium- to high-end processor with the ARMv8 ISA. The configuration we use  resembles closely real-world CPU configurations of modern critical systems from various domains, like i.e. the ARM Cortex A72 \cite{arm-a72}, used in automotive \cite{ti-jacinto}. Equally performant cores (also featuring other ISAs) are also used space-grade processors \cite{space-cpus}, like i.e. Gaisler NOEL-V \cite{gaisler}.  

\input{tables/configs}

We conduct a statistical fault injection (SFI) experiment. For each benchmark, 1000 single-bit transient (which appear only for a finite period of time) and permanent faults (stuck-at faults) are injected into the register files, randomly generated following a uniform distribution. According the widely adopted methodology from \cite{statistical_fi}, this is equal to approximate 4\% error margin with 95\% confidence level. Modelling single-bit transient and permanent faults covers sufficiently the whole range of the studied application domains, since both error types appear due to scaling \cite{CMOS_Miniaturization_Cosmic-Ray} and under heavy radiation \cite{Single_event_upset_space}, hence occur in all the reviewed use-cases.   

In our experiments, we assume that every large SRAM array (e.g., cache memories) is protected by parity or ECC schemes \cite{multi_bit_ecc, ecc_reliability_performance, mmap_ecc, riscv_ecc} and we inject faults only in the register file. The register file is a critical component in modern processors, responsible for storing and manipulating data during instruction execution. Faults in the register file can significantly affect the correctness and reliability of program execution, potentially resulting in erroneous outcomes, and impacting the entire pipeline stages. While focusing on a specific component, our evaluation indirectly assesses the overall resilience of the fault detection methods we consider, in the context of the entire pipeline. The reason is that the physical register file is not protected by any scheme due to strict performance requirements, as it is located in the critical path of the processor pipeline. This is not the case for other performance-critical hardware components, such as the load/store queue, TLBs, L1 cache memories, etc., which are usually protected, typically by a parity scheme.

\noindent\textbf{Fault Injection.} For assessing detection efficiency, we classify the outcome of each fault injection as (i) \textit{detection}: if the fault mitigation technique successfully detects and reports the error, (ii) \textit{mask}: if the execution is completed successfully and the program results were correct, (iii) \textit{silent data corruption (SDC)}: if the execution is completed successfully but produced erroneous results, (iv) \textit{crash}: if the benchmark crashed and (v) \textit{hang}: if the simulation does not terminate within three times the normal execution time.

Detection latency is measured as the elapsed time -in clock cycles- between the fault injection and the detection, while the overhead in the processor performance is evaluated with the Instructions Per Cycle (IPC) metric, in fault-free execution. 

The area overheads are calculated based on relevant literature works. Area modelling simulators, like i.e. McPAT \cite{mcpat}, provide only coarse-grained area estimations. RTL simulators on the other, can both perform fault injection and calculate more accurate area overheads, but have significantly slower simulation times, and therefore do not allow for the multiple simulation runs necessary for SFI to reach high confidence. Instead, by assessing reliability in the microarchitectural level using SFI on gem5 and basing our area analysis on previous literature, we are able to both accurately access reliability and obtain precise area measurements from real implementations. However, we do utilize McPAT to obtain power consumption estimations, since power depends highly on the workload and this would not be captured through power estimations from literature.

%% file: tables/benchmarks.tex
\begin{table}[htbp]
\caption{Benchmarks Application Domains and Classes.}
\begin{center}
\begin{tabularx}{\linewidth}{|l|X|X|}
\hline
\textbf{Benchmark} & \textbf{Application domain} & \textbf{Second-level priority class} \\ \hline
\textit{dijkstra}  & robotics \cite{route_planning_dijkstra}                                      & area/power-critical                     \\ \hline
\textit{djpeg}     & medical imaging \cite{dicom}, remote sensing                & latency-critical, area/power-critical                     \\ \hline
\textit{fft, ifft} & satellite communications, medical \cite{brigham1988fast}, HPC \cite{10.1007/11549468_136} & performance-critical                     \\ \hline
\textit{patricia}  & satellite communications, avionics             & latency-critical, area/power-critical                     \\ \hline
\textit{qsort}     & automotive, industrial control \cite{mibench}                & latency-critical, performance-critical                     \\ \hline
\textit{sha}       & IoT \cite{sha_iot}                                           & area/power-critical                    \\ \hline
\textit{edges},     & medical imaging \cite{mibench},                       & latency-critical,               \\ 
\textit{smooth}     & autonomous driving \cite{6876163} & performance-critical  \\ \hline
\end{tabularx}
\label{tab:bench}
\end{center}
\end{table}

%% file: tables/configs.tex

\begin{table}
\caption{System Configuration.}
\begin{center}
\begin{tabularx}{\linewidth}{|lX|}
\hline
\multicolumn{2}{|c|}{\textbf{Common options for all methods}} \\ \hline
CPU$^{\mathrm{a}}$ & 3-way out-of-order, 2GHz, 192-entry ROB, 64-entry IQ, 128 int + 128 float registers \\
L1 Dcache & 32kB, 8-way, 2-cycle hit latency \\
L2 cache & 2MB shared, 16-way, 20-cycle hit latency \\ \hline
\multicolumn{2}{|c|}{\textbf{R-SMT specific options}} \\ \hline
Comparison buffer size & 10 entries \\ \hline
\multicolumn{2}{|c|}{\textbf{ParDet specific options}} \\ \hline
Checker cores & 12 cores @ 1GHz, 4 stage, in-order \\
Load-store log size & 36KiB \\ \hline
\multicolumn{2}{l}{$^{\mathrm{a}}$for ParDet this refers to the main core.}
\end{tabularx}
\label{tab:config}
\end{center}
\end{table}

%% file: Evaluation.tex
\section{Evaluation}

\newcounter{mycounter}
\newcommand\insightnumber{\stepcounter{mycounter}\themycounter}

\subsection{Analysis of Detection Latency} 
\subsubsection{Re-execution slack in R-SMT} \label{sec:smt-slack}
\begin{figure}
\centerline{\includegraphics[width=\linewidth]{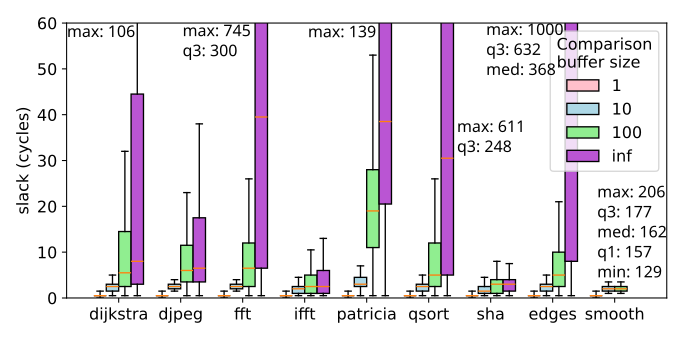}}
\caption{Distribution of re-execution slack when varying the comparison buffer size in R-SMT.}
\label{fig:slack-bufsize}
\end{figure}

For R-SMT, we define as re-execution slack (commit slack) the delay between the retirement of an instruction by the primary thread and the retirement of the same re-executed instruction by the redundant thread. To thoroughly analyze the detection latency of R-SMT, it is necessary to quantify first the re-execution slack, given it significantly impacts detection latency. In Fig. \ref{fig:slack-bufsize}, we measure the re-execution slack for various configurations, when varying the mechanism's comparison buffer size. We find that smaller buffer sizes result in smaller divergence between the primary and redundant thread during execution. This is because when the primary thread has placed more instruction results in the buffer than the redundant is able to consume, and the buffer becomes fully occupied, instruction commit of the main thread halts (named as \textit{full comparison buffer stalls}), since there are no available entries to store the subsequent instruction results. Consequently, the redundant thread is allowed to commit instead, thus converging with the primary one. Given that with smaller buffer sizes, the buffer is more frequently filled, and in turn the primary thread is more frequently denied committing, smaller comparison buffer sizes eventually lead to smaller re-execution slack.



\subsubsection{Error detection latency}
\begin{figure}
\centerline{\includegraphics[width=\linewidth]{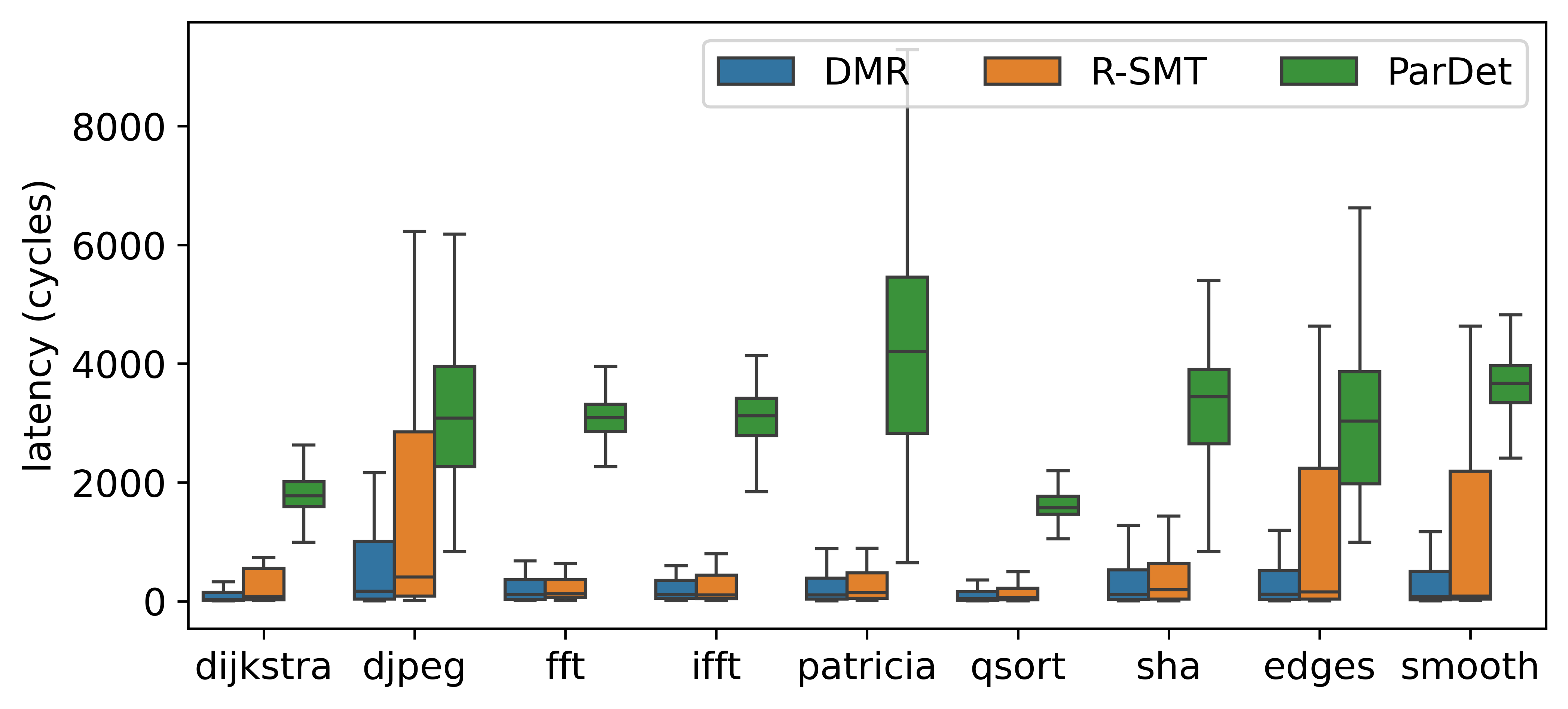}}
\caption{Distribution of detection latency across multiple injection experiments for all evaluated methods.}
\label{fig:latency}
\end{figure}

In Fig. \ref{fig:latency}, we measure the error detection latency, which yields the same trends across all benchmarks: DMR demonstrates the lowest minimum (in all benchmarks) and mean (in 8 out of 9 benchmarks) latency, because the primary and redundant instructions are concurrently executed in each core. R-SMT exhibits higher latency, because the re-execution slack introduces an \textit{additional} source of delay in the error detection, given that both the main execution and the redundant execution of an instruction must first commit before error detection can occur. Nonetheless, R-SMT demonstrates higher mean latency compared to DMR, but with close values in all benchmarks. Specifically, in 4 out of 9 (\textit{fft, ifft, qsort, smooth}) benchmarks the deviation from DMR is less than 20 cycles and for the rest, never exceeds 240 cycles (\textit{djpeg}). As shown in the previous experiment, with the given comparison buffer size (10 entries), slack is consistently lower than ten instructions, justifying the marginal latency increase observed in this experiment. ParDet consistently exhibits higher min and mean latency in all benchmarks, with up to two orders of magnitude larger compared to DMR (42x higher average for \textit{patricia}): for an error to be detected in ParDet, the whole segment needs to be first re-executed in the checker cores, which have considerably much lower compute capabilities than the main core. Moreover, checkpointing before offloading segments in a checker core, incurs additional latency overheads.

\begin{tcolorbox}
\noindent\textbf{Key insight \insightnumber:} For R-SMT, slack between threads is the key factor that causes the high detection latency.
\end{tcolorbox}

\begin{tcolorbox}
\noindent\textbf{Key insight \insightnumber:} Although ParDet might have the potential to detect errors early, since it performs architectural state comparisons in re-executions, these comparisons happen infrequently due to the low compute capabilities of the checker cores, thus detection latency is consistently larger than DMR and R-SMT.
\end{tcolorbox}


\subsection{Analysis of Detection Efficiency}

\begin{figure}
\centerline{\includegraphics[width=\linewidth]{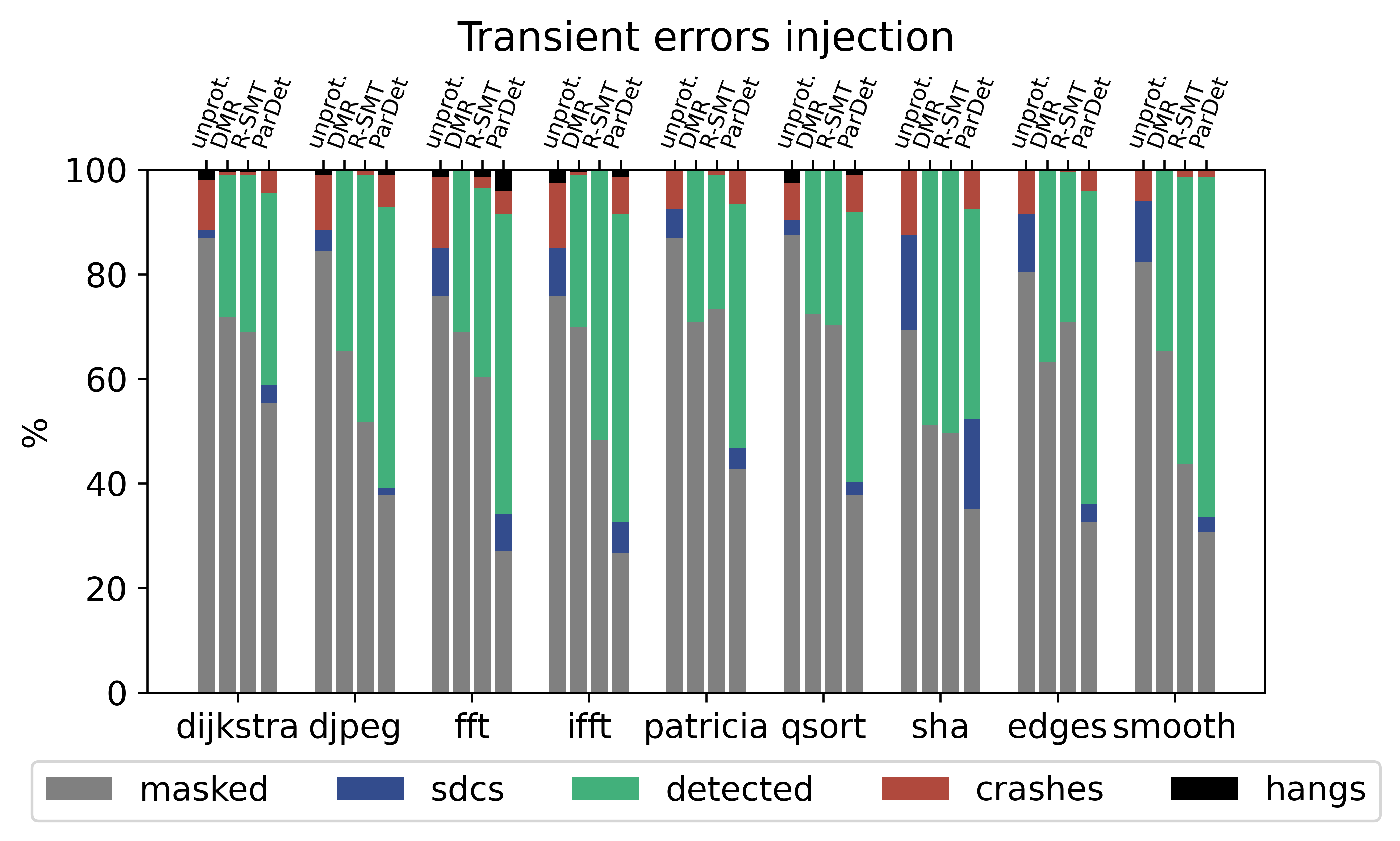}}
\caption{Detection efficiency in transient errors.}
\label{fig:transient}
\end{figure}

To assess the effectiveness of error detection in each method, we also inject errors to an unprotected system to use as a baseline.

Fig. \ref{fig:transient} compares the error detection efficiency in the presence of transient faults.  We observe that ParDet detects the largest number of errors across all benchmarks (52\% on average, compared to 39\% for R-SMT and 34\% for DMR). This is because ParDet's implementation of architectural state comparison for error detection, which detects cases where the injected erroneous register will not affect any instruction/the execution, i.e. when the erroneous register is overwritten. In contrast, DMR and R-SMT correctly classify such cases as masked, since no erroneous result is produced. Masked injections occur both in the unprotected and protected designs, because the erroneous register is overwritten, given that the fault injection can occur in unmapped registers (it depends on the fault masks generation, which follows the uniform distribution according to \cite{statistical_fi}), so the value of that register will never be used.

\begin{tcolorbox}
\noindent\textbf{Key insight \insightnumber:} High detection percentage does not guarantee low crash percentage, as methods such as ParDet might suffer from \textit{overdetection}. 
\end{tcolorbox}

Overdetection, where benign errors that will be subsequently masked are identified as detected, creates false positives, which is generally unappealing. For instance, in systems where errors are also corrected after detection, false positives lead to unnecessary corrective actions, which introduce additional performance overhead and increase response time.

R-SMT detects more faults compared to DMR, owned to the structure of the injection campaign: to fairly compare fault detection capabilities, the same errors are injected in all methods, and to better approximate the response of fault detection in real-world execution scenarios, each injection is \textit{synchronized} to occur for all methods when the \textit{same} time interval has elapsed (sampled from a normal distribution and measured from the start of the program execution). This results in errors manifesting earlier in the program order in R-SMT, thus causing a higher probability that the error will propagate to other registers via data dependencies. Specifically, R-SMT occurs a higher probability to detect errors, as errors potentially corrupt more instructions.

Despite DMR exhibiting the lowest detection percentage, it experiences the fewest failures (0.1\% on average, crashes and hangs combined), followed by R-SMT (1.1\%) and then ParDet (6.2\%). This is because R-SMT and ParDet have higher detection latency than DMR, which might cause the error to propagate deeper, and in turn causing failures (crashes and hangs) more frequently than DMR.


Lastly, neither DMR nor R-SMT produces SDCs, because for an erroneous value to be written in program output, it must first be produced in some instruction result and consequently, would have been detected.



\begin{figure}
\centerline{\includegraphics[width=\linewidth]{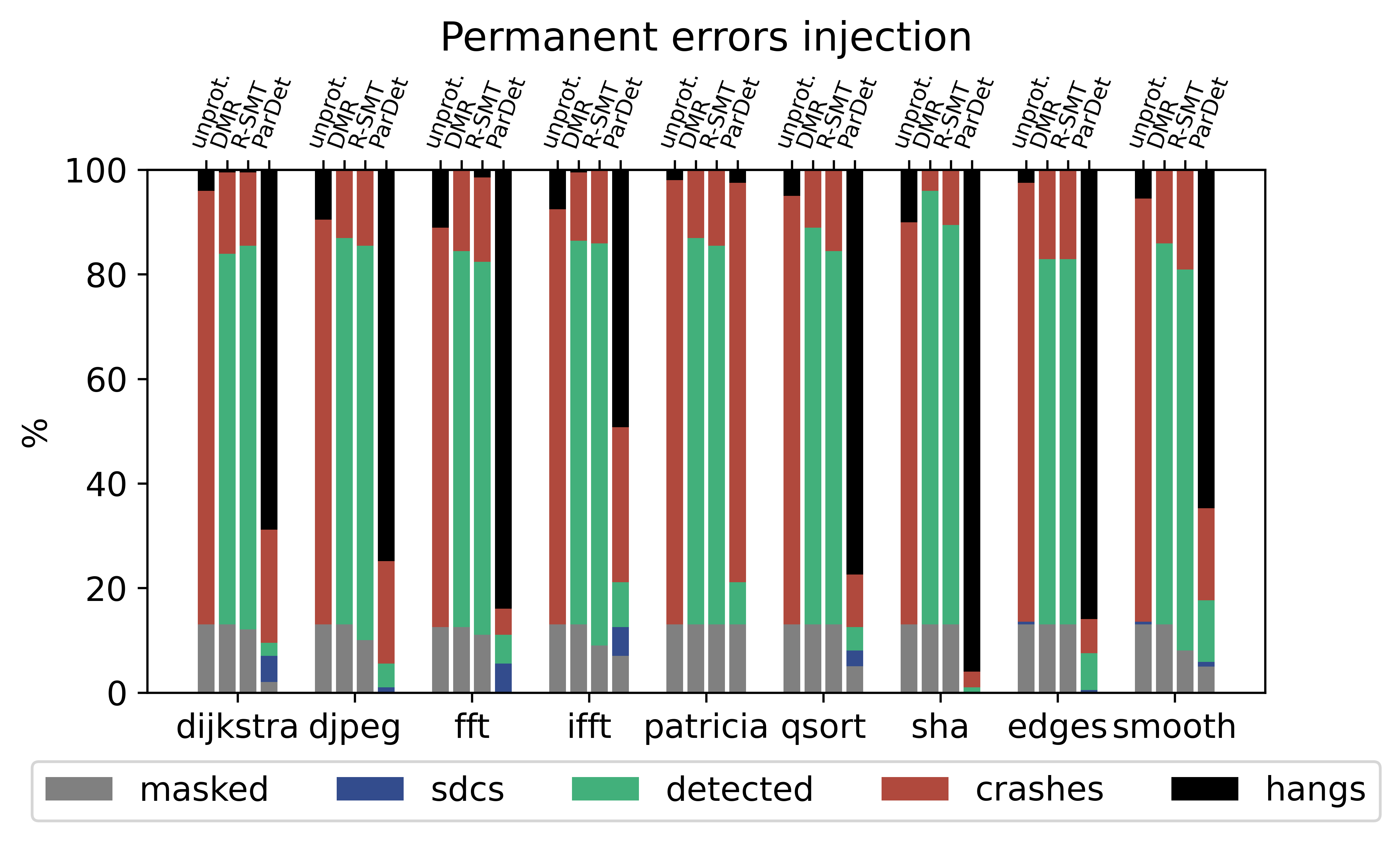}}
\caption{Detection efficiency in permanent errors.}
\label{fig:permanent}
\end{figure}

Fig. \ref{fig:permanent} compares the efficiency in permanent error detection. All methods experience more crashes and hangs, since permanent errors propagate more, due to their persistent nature, i.e., affecting more instructions. This, combined with the increased detection latency of ParDet renders it the method with the lowest permanent error detection efficiency. In contrast, DMR and R-SMT demonstrate higher detection efficiency, because permanent errors through their wider propagation more likely  corrupt instruction results, thus making them detectable by these methods.

\begin{tcolorbox}
\noindent\textbf{Key insight \insightnumber:} R-SMT despite duplicating hardware, can effectively detect permanent errors, by detecting altered instruction results before and after the permanent hardware corruption.
\end{tcolorbox}

\begin{tcolorbox}
\noindent\textbf{Key insight \insightnumber:} Low detection latency is even more crucial in permanent error detection, since it provides higher detection efficiency.
\end{tcolorbox}


\subsection{Performance Analysis}

\begin{figure}
\centerline{\includegraphics[width=\linewidth]{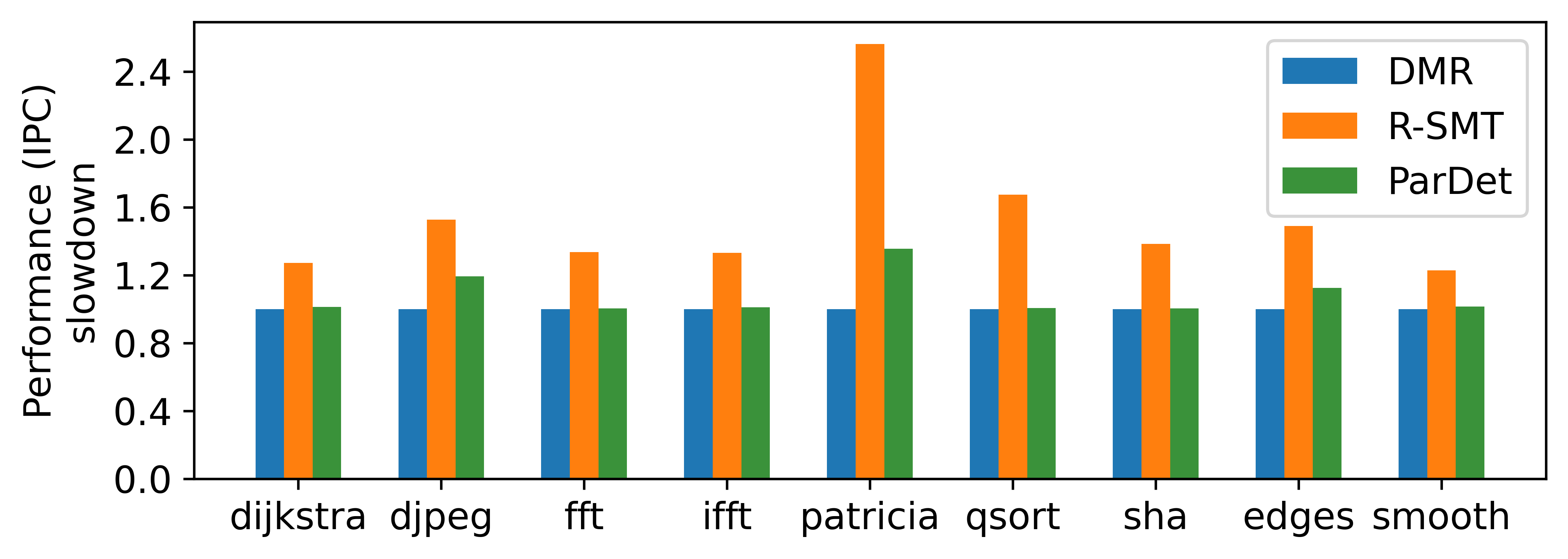}}
\caption{Comparison of performance in error-free execution for all 3 methods.}
\label{fig:ipc}
\end{figure}

\begin{figure}
\centerline{\includegraphics[width=\linewidth]{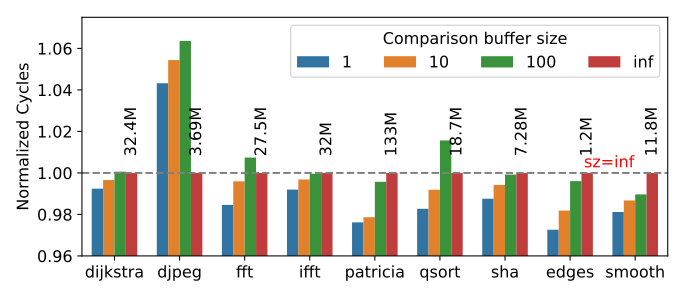}}
\caption{Slowdown of R-SMT in error-free execution, when varying the size of the comparison buffer.}
\label{fig:cycles_vs_bufsize}
\end{figure}

Fig. \ref{fig:ipc} presents the performance overhead each method introduces to the execution using the IPC metric. DMR emerges as the most performant across the three, providing on average 1.53x and 1.08x better peformance than R-SMT and ParDet, respectively, as it introduces no performance degradation from normal (unprotected) execution. ParDet experiences a performance slowdown less than 2\% in all benchmarks except from \textit{djpeg}, \textit{patricia} and \textit{edges}: checkpointing causes more frequent stalls of the main core in these benchmarks. In all benchmarks, ParDet is more performant than R-SMT by 1.4 times on average.
Lastly, the performance overhead of R-SMT is due to two reasons: i) the performance degradation of SMT (2-thread execution) compared to single threaded execution, and ii) the performance loss due to full comparison buffer stalls. Fig. \ref{fig:cycles_vs_bufsize} shows the impact of full comparison buffer stalls of R-SMT scheme for various comparison buffer sizes. We find that larger sizes improve performance by causing more infrequent full comparison buffer stalls of the primary thread. 


\subsection{Area Cost Analysis}

In Table \ref{tab:area} we compare the area costs of all three methods over the unprotected baseline that does not integrate fault detection schemes. For DMR, we assume a two-times area overhead, since all hardware components of the core are replicated. Given that area modelling simulators, i.e. McPat \cite{mcpat}, do not estimate the area overhead of SMT, we evaluate the area overheads of R-SMT by relevant literature \cite{marr_hyper-threading_2002, koufaty_hyperthreading_2003, preston_design_2002}, in order to also obtain more accurate results: the layout overhead of SMT in real designs is less than 6\% of the core's area, excluding cache memories, and the comparison buffer of 10 entries corresponds to 0.125\% of the L1 cache thus contributing to an additional 0.04\% area increase \cite{Ramon_2015}. For ParDet, we use the results from the original publication \cite{ainsworth} (since we are also modelling the same microarchitecture), i.e., incurring 24\% (0.14mm$^2$/core and 0.08mm$^2$ for the 36KiB load-store log) overhead over the main core. 

\input{tables/area}

\begin{tcolorbox}
\noindent\textbf{Key insight \insightnumber:} R-SMT configurations which maintain small slack will harness multiple benefits: without needing a large comparison buffer, will enjoy both low area overhead and better performance, due to infrequent stalls.
\end{tcolorbox}

\subsection{Power overheads analysis}

\begin{figure}
\centerline{\includegraphics[width=\linewidth]{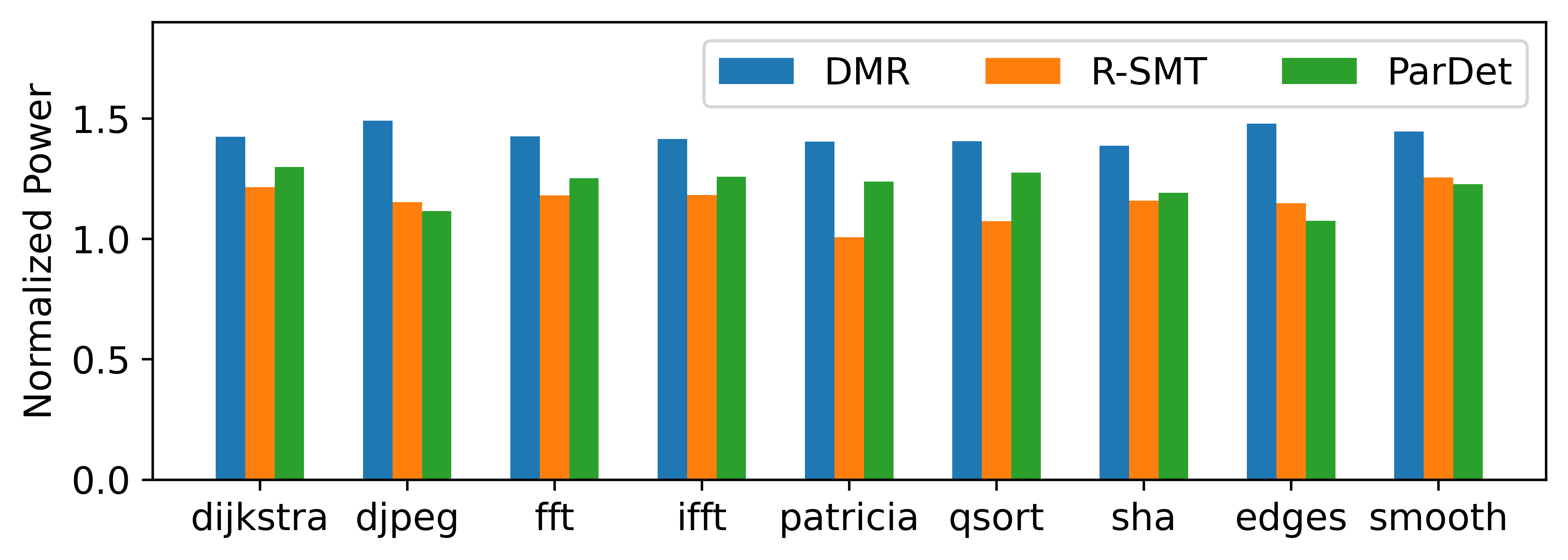}}
\caption{Power consumption for all 3 methods, normalized to the unprotected design.}
\label{fig:power}
\end{figure}

Fig. \ref{fig:power} shows the total power consumption of all methods across all benchmarks, estimated using McPAT and normalized to the power consumption of the unprotected design, which consists of a single-core processor. DMR exhibits on average power overhead of 43\%, given that introduces an additional core and duplicate instructions. In contrast, R-SMT demonstrates the lowest power increase among all methods, with only 15\% increase over the unprotected baseline. This minimal increase is due to maintaining the same core configuration for all methods, with the power increment attributed only to the execution of the redundant SMT thread.

If the processor microarchitecture were to be more aggressive to support SMT initially (e.g., with larger ROB or register files, etc.), a further increase in power consumption would be anticipated. However, since the redundant thread does not significantly congest the main thread due to re-execution slack, we retained the same microarchitecture for R-SMT as well, not augmenting it with more resources for SMT.

Lastly, ParDet incurs a 21\% power overhead, due to the additional power consumption of the checker cores.  

Since McPAT provides only coarse estimations for area and power, we expect more accurate estimations of power to yield even lower results, when using more accurate and recent models tailored to current technology.  

\subsection{Overall Evaluation}
\input{tables/tradeoffs}

In Table \ref{tab:tradeoffs}, we evaluate various configurations of R-SMT and ParDet, each having a different area cost, using all metrics and present averaged numbers across all benchmarks. For R-SMT, we modify the size of the comparison buffer. For ParDet, we adjust the number of checker cores, scaling the load-store log such that the size that corresponds to each core remains constant in every configuration, since otherwise detection latency would be affected primary by that. 

ParDet with 3 auxiliary cores and all configurations of R-SMT exhibit similar area overheads. However, due to the low core count in checker cores, ParDet achieves lower performance over R-SMT by nearly a factor of two (2.92x slowdown compared to 1.53x) and higher error detection latency by a factor of 20 (96-149 cycles compared to 3011 cycles). Additionally, ParDet experiences 6 times more crashes, with the increased detection percentage attributed to overdetection (detecting also benign errors which would be eventually masked). Thus, in terms of area cost, R-SMT provides a good trade-off across all metrics.

\begin{tcolorbox}
\noindent\textbf{Key insight \insightnumber:} R-SMT is more suitable for area-critical applications.
\end{tcolorbox}

In ParDet configurations, increasing the core count reduces the detection efficiency, because with more cores, the main one progresses further, and errors propagate further without being limited when the main core stalls. However, crashes do not increase, because checkpoints continue to occur at the same frequency, as the segment size per core remains constant.

\begin{tcolorbox}
\noindent\textbf{Key insight \insightnumber:} ParDet presents a tradeoff between performance and detection efficiency versus latency and area.  
\end{tcolorbox}

R-SMT configurations with varying comparison buffer size, perform similarly in detection efficiency and detection latency. This is because error detection continues to occur as early as possible during the redundant thread's commit phase.

\begin{tcolorbox}
\noindent\textbf{Key insight \insightnumber:} Both R-SMT’s detection efficiency and latency are independent of area configuration (comparison buffer size), thus hardware designers can safely select the lowest-area configuration without compromising neither.
\end{tcolorbox} 

\begin{tcolorbox}
\noindent\textbf{Key insight \insightnumber:} ParDet with increased core counts is suitable for performance-critical applications, since along with adequate detection efficiency has minimal performance overheads. 
\end{tcolorbox}

\begin{tcolorbox}
\noindent\textbf{Key insight \insightnumber:} DMR or R-SMT with the lowest comparison buffer sizes are equally good options for latency-critical applications, since both provide the minimal latency. 
\end{tcolorbox}

%% file: tables/area.tex
\begin{table}[htbp]
\caption{Area Overheads.}
\begin{center}
\begin{tabular}{|c|cc|c|}
\hline
\multirow{2}{*}{\textbf{Method}} & \multicolumn{2}{c|}{\textbf{Overhead per Component}} & \multirow{2}{*}{\textbf{\begin{tabular}[c]{@{}c@{}}Total \\ area\end{tabular}}} \\ \cline{2-3}
 & \multicolumn{1}{c|}{\textit{\textbf{Component}}} & \textit{\textbf{Overhead}} &  \\ \hline
Unprotected & \multicolumn{1}{c|}{-} & - & 1x \\ \hline
DMR & \multicolumn{1}{c|}{Redundant core} & 100\% & 2x \\ \hline
\multirow{2}{*}{R-SMT} & \multicolumn{1}{c|}{SMT overhead} & 6\% & \multirow{2}{*}{1.0604x} \\ \cline{2-3}
 & \multicolumn{1}{c|}{\begin{tabular}[c]{@{}c@{}}Comparison buffer (10 entries)\end{tabular}} & 0.04\% &  \\ \hline
\multirow{2}{*}{ParDet} & \multicolumn{1}{c|}{Checker cores (12)} & 20.2\% & \multirow{2}{*}{1.24x} \\ \cline{2-3}
 & \multicolumn{1}{c|}{Load-store log (36 KiB)} & 3.8\% &  \\ \hline
\end{tabular}
\label{tab:area}
\end{center}
\end{table}

%% file: tables/tradeoffs.tex
\begin{table*}[t]
\caption{Overall evaluation for various configurations}
\begin{center}
\begin{tabular}{|cl|l|l|l|l|l|l|l|l|}
\hline
\multicolumn{2}{|c|}{\textbf{Method}} & \multicolumn{1}{c|}{\textbf{R-SMT}} & \multicolumn{1}{c|}{\textbf{R-SMT}} & \multicolumn{1}{c|}{\textbf{R-SMT}} & \multicolumn{1}{c|}{\textbf{ParDet}} & \multicolumn{1}{c|}{\textbf{ParDet}} & \multicolumn{1}{c|}{\textbf{ParDet}} & \multicolumn{1}{c|}{\textbf{ParDet}} & \multicolumn{1}{c|}{\textbf{DMR}} \\ \hline
\multicolumn{2}{|c|}{Configuration} & 1-entry & 10-entry & 100-entry & 3 cores & 6 cores & 12 cores & 16 cores & - \\ \hline
\multicolumn{2}{|c|}{Area overhead} & 6.004\% & 6.04\% & 6.4\% & 6\% & 12\% & 24\% & 32\% & 100\% \\ \hline
\multirow{5}{*}{Detectability} & detected \% & 37.7 & 38.9 & 33.7 & 53.8 & 52.6 & 52.2 & 51.9 & 34.7 \\
 & SDC \% & 0 & 0 & 0 & 5.2 & 5.9 & 5.4 & 5.6 & 0 \\
 & masked \% & 62.2 & 60 & 63.2 & 35.1 & 35.6 & 36.2 & 36 & 65.3 \\
 & crashes \% & 0.1 & 0.9 & 1 & 5.9 & 5.5 & 5.4 & 4.2 & 0 \\
 & hangs \% & 0 & 0.2 & 2.1 & 0 & 0.4 & 0.8 & 2.3 & 0 \\ \hline
\multicolumn{2}{|c|}{Performance slowdown} & 1.54 & 1.53 & 1.52 & 2.92 & 1.56 & 1.08 & 1.06 & 1 \\ \hline
\multicolumn{2}{|c|}{Latency (cycles)} & 96 & 149 & 108 & 3011 & 3133 & 3136 & 3149 & 96 \\ \hline
\end{tabular}
\end{center}
\label{tab:tradeoffs}
\end{table*}

%% file: OtherRelated.tex
\section{Other Related Work}
\subsection{Error Detection at Other Architecture Levels}
A couple of prior works~\cite{reis_design_2005,gizopoulos_online_error_detection} propose hybrid software-hardware and software-only error detection methods. In this work we focus on a comprehensive analysis of only hardware error detection methods, as opposed to i.e. software-level detection. This focus is crucial, because the selection of any hardware method impacts heavily the late stages (where the impact on area and power is more accurately computed) albeit must be made during the early design stages and cannot be altered afterwards. Consequently, the insights we present impact significantly all stages of hardware development. 
Additionally, error detection is a necessity in modern computing systems, thus manufacturers have widely integrated error detection schemes in hardware. Therefore, our work focuses in effectively evaluating hardware-level methods that have higher detection efficiency and typically better performance than software-based schemes.

\subsection{Error Correction}
Several prior works~\cite{vijaykumar_transient-fault_2002, paramedic,mukherjee_cache_2004,hagbae_kim_evaluation_2000,wunderlich_efficacy_2013,Energy_Efficiency_Protocols} propose error correction methods, i.e., schemes to effectively resolve errors after they have been detected. Error correction is orthogonal to error detection: a self correcting system has to first detect errors prior to correcting them, and the overhead of correction typically dominates that of detection, as restoring the system to a safe state requires additional processing \cite{vijaykumar_transient-fault_2002, paramedic}. For this reason, a comparison between both correction and detection would not be fair. Instead, we leave the characterization of error correction methods for future work, as they will potentially require different metrics than the ones proposed here, again depending on the requirements of the applications utilizing correction.

\subsection{Other Characterization Studies}

Prior characterization studies between hardware methods (for detection or also correction) include only limited metrics. One study \cite{mukherjee_cache_2004} assesses scrubbing, a hardware correction technique for memory faults, in terms of reliability. \cite{hagbae_kim_evaluation_2000} compares the latency of hardware redundancy schemes for error correction, \cite{wunderlich_efficacy_2013} compares the reliability of detection techniques for GPUs and \cite{Energy_Efficiency_Protocols} discusses the energy efficiency of fault tolerant methods for parallel HPC systems. 
Reis et al. \cite{reis_design_2005} propose a hybrid software-hardware method evaluated in performance, reliability, and area in comparison to a software-only method. Gizopoulos et al. \cite{gizopoulos_online_error_detection} discuss hardware and software methods qualitatively comparing them in performance, reliability, detection latency and area overhead, but without evaluating them experimentally in neither of these metrics. In this work we focus on multi-metric evaluation of hardware-level methods, covering all the important metrics needed across all application domains.


%% file: Conclusion.tex
\section{Conclusion}
We presented a comprehensive characterization of three diverse alternatives on hardware error detection. We identified that safety-critical applications from different domains can be classified in three categories based on their second requirement beyond reliability. We proposed five key metrics on which fault detection methods need to be evaluated to effectively cover all various application classes. We extensively compared three methods, i.e., (i) DMR a conventional fault tolerance method, and (ii) R-SMT and (iii) ParDet two microarchitectural approaches, in all the proposed metrics using various workloads. Our work demonstrates that microarchitectural methods achieve comparable detection efficiency with the conventional one, and proposes that (i) R-SMT can be used in area/power-critical applications, since it provides low area and power without compromising detection efficiency in latency, (ii) both R-SMT and DMR are suitable for latency-critical applications and (iii) ParDet fits very well for performance-critical applications.